# Accretion rates and radiative efficiencies of Sagittarius A* and nearby supermassive black holes estimated using empirical relations: Implications for accretion models

Yash Aggarwal

Emeritus Associate: Lamont-Doherty Earth Observatory
Palisades, NY 10965, USA
Current address: 822 Winton Drive, Petaluma, CA 94954
E-mail: haggarwal@hotmail.com

## ABSTRACT

The Bondi accretion rate $\dot{M}_B$ of black holes (BHs) in our and nearby galaxies Messier 87, NGC 3115, NGC 1600, and Cygnus A have been determined/constrained using Chandra or other observations It, however, remains unknown how much mass from the Bondi radius reaches each BH and how much is accreted. We determine the accretion rate $\dot{M}$ and radiative efficiency $\varepsilon$ for each BH using two well-tested empirical relations: one relates a BH's mass-accretion rate $\dot{M}$ to its mass and redshift, and the other relates $\varepsilon$ to Eddington ratio and redshift. We get $\dot{M}=2\times10^{-5}$ solar masses/year and $\varepsilon\sim0.9$ for Sagittarius A*, $\dot{M}\sim0.09$ solar masses/year and $\varepsilon\sim0.68$ for NGC 1600; and values in between these extremes for the rest. The derived mass-inflow rate onto each BH essentially matches the reported $\dot{M}_B$. Thus, the results do not support the ADIOS and CDAF models, but whether the dissipated energy not reflected in a BH's observed luminosity is advected as in the ADAF model remains uncertain. Furthermore, contrary to current model expectations, the derived $\varepsilon$ are orders of magnitude higher and $\varepsilon$ increases as accretion rate decreases and a BH ages. A physical basis is found relating empirical formulation of $\dot{M}$ to Bondi accretion.

# 1. INTRODUCTION

Supermassive black holes (SMBHs) can be grouped into two broad classes based on their observed luminosities: those mostly at high redshifts accreting at relatively high rates whose observed bolometric luminosities are a substantial fraction of their Eddington luminosities; and those in nearby galaxies accreting at very low rates whose observed luminosities are orders of magnitude lower than their Eddington luminosities (see Yuan and Narayan, 2014; henceforth YN14). The active galactic nuclei (AGNs) in the latter class are often referred to as low-luminosity or radiatively inefficient AGNs. Their observed luminosity deficit is attributed to hot accretion flows described by three variants. In the advection-dominated accretion flow (ADAF; Narayan and Yi, 1994; Abramowicz et al. 1995; Chen et al. 1995) the mass-inflow rate from the Bondi radius to the black hole (BH) is constant, but most of the dissipated accretion energy is advected instead of being radiated away. In the adiabatic inflow-outflow solution (ADIOS; Blandford & Begelman, 1999; 2004) the mass inflow rate decreases towards the BH. And in the convection dominated accretion flow (CDAF; Narayan, Igumenshchev, and Abramowicz, 2000; Quataert & Gruzinov, 2000) gas is locked in convective eddies and the inflow rate decreases towards the BH. The three models predict different density ($\rho$) profiles

Sagittarius A* (Sgr A*) is a prime example of the LLAGN class and the first SMBH whose Bondi accretion rate was estimated using observational data. Quataert, Narayan & Reid (1999; henceforth QNR1999) used observed distribution of mass-losing stars and assuming non-interacting winds; and Cocker and Melia (1997; henceforth CM1997) used Bondi-Hoyle 3D hydro dynamical numerical simulations of accretion from winds emanating from nearby massive stars. In addition, Sgr A* has been extensively studied using accretion models and analyses of its radio emissions. Depending upon the assumptions used, the models suggest that its accretion rate $\dot{M}$ is lower than its $\dot{M}_B$ by 2-4 orders of magnitude. And combined with its extremely low observed luminosity, these results have led to the inference that its radiative efficiency must be extremely low and that most of the mass available at the Bondi radius does not reach Sgr A* ( see review by YN 14 and references therein). There are four additional SMBHS whose $\dot{M}_B$ have now been determined or constrained. Three of the determinations are based on measured density and temperature profiles away. These are SMBHs in nearby galaxies Messier 87 (Di Matteo et al., 2003; Russell et al., 2015 and 2018; henceforth R15 and R18), NGC 3115 (Wong et al., 2011; Wong et al., 2014; henceforth W11 and W14), and NGC 1600 (Runge and Walker, 2021; henceforth RW21). The fourth is Cygnus A whose $\dot{M}_B$ was estimated by Lo et al. (2021) using polarimetric observations for different density profiles. The observed density profiles, however, do not conclusively discriminate between the three accretion models, and the questions as to what fraction of the mass available at the Bondi radius reaches the BH and how much of it is accreted or radiated away by the BH remain unresolved.

This letter attempts to answer the foregoing questions by estimating the accretion rate and radiative efficiency of each of the five BHs using two empirical relations. The empirical relations, described hereunder, were derived by Aggarwal (2023, 2022; henceforth A23 and A22) using available mass and redshift data of the highest-z SMBHs. They are essentially free of assumptions and have been extensively tested. The primary empirical relation defines a BH's accretion (not inflow) rate $\dot{M}$ in units of solar mass/year ($M_\odot$/yr) as a function of its mass $M_{BH}$ and redshift z.

$$\dot{M} (M_\odot/\text{yr}) = 4.96 \times 10^{-12} M_{BH} (1 + z)^3 \qquad (1)$$

A secondary empirical relation defines a BH's Eddington ratio $\lambda$ as a function of its redshift z and radiative efficiency $\varepsilon$.

$$\lambda = 2.18 \times 10^{-3} (1 + z)^3 \varepsilon /(1- \varepsilon) \qquad (2)$$

Equation 2 is derived using Eq.1 and the definition of $\lambda = L_{bol}/L_{Edd}$, where $L_{bol}$ is the bolometric luminosity and Eddington luminosity $L_{Edd} = 1.3 \times 10^{38} M_{BH}$ in ergs/s and $M_\odot$ in solar masses. Radiative efficiency $\varepsilon$ is conventionally defined with respect to the mass inflow rate $\dot{M}_i$, and a BH's accretion rate $\dot{M}$ is smaller by a factor of $(1 - \varepsilon)$. Hence, $L_{bol} = (\dot{M}c^2) \varepsilon / (1- \varepsilon)$, where c is the velocity of light. Note that a BH's radiative efficiency $\varepsilon$ can be obtained using Eq.2 if its Eddington ratio $\lambda$ is known.

Determining a BH's spot mass accretion rate $\dot{M}$ is rather straight forward using Eq.1. All we need to know is a BH's mass $M_{BH}$ and redshift. Translating $\dot{M}$ into the mass inflow rate $\dot{M}_i$ onto the BH, however, requires knowledge of a BH's radiative efficiency $\varepsilon$ which can be estimated using Eq.2 if a BH's Eddington ratio $\lambda$ is known. The Eddington ratios of all but the BH in Cygnus A are, however, not known. We estimated their $\varepsilon$ using Eq. 2, available data for $\lambda$ for similar-size BHs at similar redshifts, and observed dependence of $\lambda$ on a BH's mass and redshift found by A23. The methodology used will become clear later when applied. Having thus estimated $\dot{M}$ and $\varepsilon$, we compare the resulting mass inflow rate $\dot{M}i$ with each BH's Bondi accretion rate $\dot{M}_B$, discuss the implications of the findings, seek a physical basis relating Eq.1 to the theoretical equation defining the Bondi accretion rate, and end with conclusions.

## 2. RESULTS

To calculate the predicted $\dot{M}$ using Eq.1, we used the same $M_{BH}$ or the range of $M_{BH}$ as used in the studies that determined $\dot{M}_B$, except for Sagittarius A* for which we used the current best estimate of $4.15 \times 10^6 M_\odot$ (Gravity Collaboration, 2019) instead of the best estimate of $2.6 \times 10^6 M_\odot$ then available and used by QNR1999 and CM1997 and others. As pointed out earlier, radiative efficiency $\varepsilon$ can be estimated using Eq.2, but the Eddington ratios $\lambda$ of all but one of the 5 BHs are unknown. A23 found that $\lambda$ is a func6tion of a BH's $M_{BH}$ and redshift as demonstrated graphically in Fig.1. It shows a spline regression plot of $\lambda$ versus z based on Kozlowski (2017) catalog of 132,000 BHs separated into 3 mass bins. Hence, we sought to find in the literature BHs with known $\lambda$ values closest in mass and redshift to each one of our 5 BHs; and used the average of the $\lambda$ values found to estimate the radiative efficiency using Eq.2. The $M_{BH}$ and redshifts of the five BHs are listed in Table1. The spline regression plot in Fig.1 shows that $\lambda$ decreases with z, but reaches a near constant at z< 0.5 for BHs in the green and blue bins. Of the five BHs, M87, NGC 3115, and Cygnus A have masses that are comparable to those in the green bin in Fig.1 and have redshifts z <0.06. Kozlowski lists 35 quasars at z<0.5 in the green bin within a narrow z window of ~ 0.492-0.345 having a mean $\lambda$=0.0118. Adopting this value of $\lambda$ for the 3 BHs and substituting it in Eq.2 along with each BH's redshift, we get $\varepsilon$ =0.84 for M87 and NGC 3115 and $\varepsilon$ =0.82 for Cygnus A. Independently, Vasudevan et al. (2010) determined $\lambda$=0.017 for Cygnus A. Using this value of $\lambda$ for Cygnus A in Eq.2, one gets $\varepsilon$

=0.867, which is comparable to that (0.82) estimated using Fig.1 and Kozlowski's λ data for z<0.5. NGC 1600 is the most massive of the five BHs and it falls in the blue bin in Fig.1. The data in Fig.1 show that larger BHs have lower λ than smaller ones. Hence NGC 1600 with z comparable to those of M87 and NGC 3115 should have a lower λ than ~0.84. The difference in the λ values for the green and blue bins at z<0.5 is, however, not discernable in Fig1 because of the scale used. But Kozlowski's catalog shows that the average λ for the BHs in the blue bin decreases from ~0.05 at z=1.5 to ~0.034 near z=1, to ~0.0075 near z=0.73, and ~ 0.0049 at z=0.36. Hence, using λ=0.0049 and z of NGC 1600 in Eq.2, we get ε=0.68 for NGC 1600. Sgr A* is smaller than the other four by 3-4 orders of magnitude. Hence, its λ should be higher than 0.84. Vasudevan et al. (2010) list 4 BHs at 0.002 <z < 0.006 with an average $M_{BH}$ ~ $5.7 \times 10^6$ $M_\odot$ close to that of Sgr A* and an average λ =0.019. Using λ =0.019 in Eq.2, we get ε=0.90 for Sgr A*. It is noteworthy that these estimates of ε are consistent with the finding by A23 that ε is an inverse function of BH mass $M_{BH}$. Of the 5 BHs, NGC 1600 is the largest and its ε is the lowest, and Sgr A* is the smallest and its ε is the highest. The rest have $M_{BH}$ and ε in between these two extremes. The uncertainty in ε arising from the uncertainty in λ is apparently relatively small as indicated by the following observations. The adopted λ for Sgr A* is roughly twice that of M87 and NGC 3115, but its ε is higher by only ~10% despite the fact that its $M_{BH}$ is ~3 orders of magnitude lower. Similarly, the λ of NGC 1600 is ~ a factor of 2 lower than that of M87, NGC 3115, and Cygnus A, but its ε is lower by only ~19%. These observations indicate that the change in ε due to an uncertainty in λ is relatively small, and hence the ε of the 5 BHs inferred using Eq.2 are probably rather well constrained.

The results are shown in Table 1. In addition to a BH's mass $M_{BH}$ and redshift z, Table 1 lists the inferred radiative efficiency ε, the derived mass accretion rate $\dot{M}$, the derived mass inflow rate $\dot{M}_i$, the reported Bondi accretion rate $\dot{M}_B$, and the derived bolometric luminosity $L_{bol}$ for each BH. The mass inflow rate $\dot{M}_i$ on to each BH is simply $\dot{M}$ divided by (1- ε) and $L_{bol}$ = ($\dot{M}c^2$) ε / (1- ε), where c is the velocity of light. For a given redshift, $\dot{M}$ scales as $M_{BH}$ and the uncertainty in $\dot{M}$ is directly proportional to the uncertainty in $M_{BH}$ (see Eq.1). And since ε is rather well constrained as discussed above, the uncertainty in $\dot{M}_i$ is also largely proportional to the uncertainty in $M_{BH}$. On the other hand, $\dot{M}_B \propto c_s^{-1} \rho_B R_B M_{BH}$, where $c_s$ and $\rho_B$ are the velocity of sound and gas density at the Bondi radius $R_B$ (Bondi, 1952). For a given density and temperature profile, $\dot{M}_B$ also appears to roughly scale as $M_{BH}$ as suggested by the following observations. The lowest and the highest $M_{BH}$ of M87 differ by a factor of ~2 and so do their range of reported $\dot{M}_B$. The same is exactly the case for NGC 3115. And for NGC 1600, the highest and the lowest $M_{BH}$ differ by a factor of ~1.8 and the highest and lowest reported $\dot{M}_B$ differ by a factor of 2. Thus, $\dot{M}_i$ and $\dot{M}_B$ are apparently equally affected by the uncertainty in $M_{BH}$. Lastly, note also that previous estimates of $L_{bol}$ for these BHs are based on their $\dot{M}_B$ and assume a fixed value of 0.1 for radiative efficiency, whereas the $L_{bol}$ estimated here are based on each BH's predicted $\dot{M}$ and derived radiative efficiency. The predicted $L_{bol}$ are on the average higher by a factor of ~8 than previous estimates using ε=0.1 except for Cygnus A.

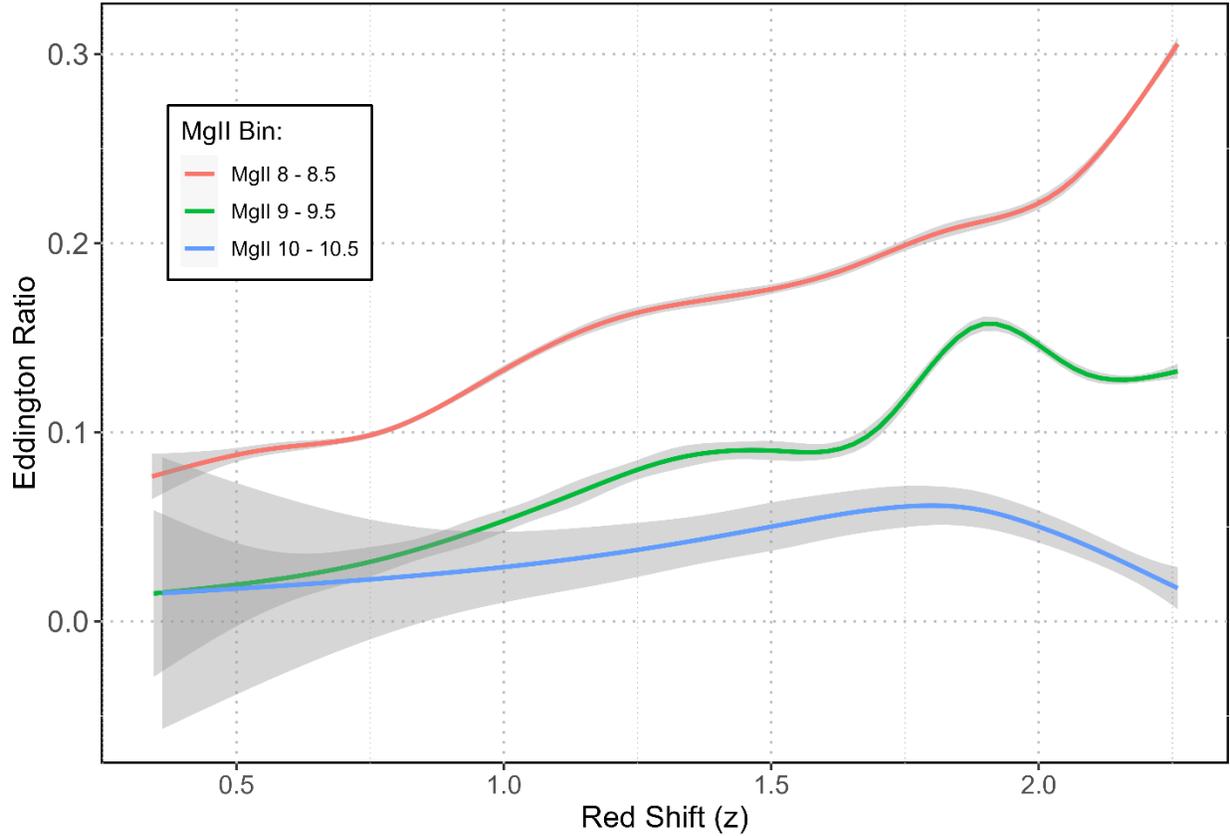

Fig.1: Spline regression plot of Eddington ratio $\lambda$ versus z using ggplot2 by Wickham (2016). The data are from Kozlowski's (2017) catalog separated into 3 mass $M_{BH}$ bins. The $1-3 \times 10^8 \, M_\odot$ red group has 36,871; the $1-3 \times 10^9 \, M_\odot$ green group has 28,799; and the $1-3 \times 10^{10} \, M_\odot$ blue group has 523 AGNs. Each line gives the median value of $\lambda$ as a function of z, and the grey area around it shows the 95% confidence interval. A fourth $M_{BH}$ group of $1-3 \times 10^7 \, M_\odot$ with 674 BHs (not plotted because of scale differences) in fact shows a more pronounced decrease in $\lambda$ with z. (From A23).

## 3 DISCUSSION

Let us first examine the results for M87, NGC 3115, and NGC 1600 for which the reported $\dot{M}_B$ are based on measured density and temperature at or near their Bondi radii, and hence are the most reliable. Recall that the range of values shown for $\dot{M}_i$ and $\dot{M}_B$ for each of the 3 BHs correspond to the range of $M_{BH}$ values used to compute $\dot{M}_i$ and $\dot{M}_B$. For M87, the range of predicted $\dot{M}_i$ values is essentially identical to the range of reported $\dot{M}_B$ values. For NGC 3115, the lower $\dot{M}_i$ value is in between the reported $\dot{M}_B$ values, and the higher $\dot{M}_i$ is within a factor of ~1.5 of the higher $\dot{M}_B$. For NGC 1600, $\dot{M}_i$ is within a factor of <2 of $\dot{M}_B$. We note, however, that $\dot{M}_B$ of NGC 1600 may be significantly underestimated for the following reason. As indicated above, both $\dot{M}_i$ and $\dot{M}_B$ scale as $M_{BH}$. The $M_{BH}$ of M87 is 3-4 times higher than that of NGC 3115, and its $\dot{M}_B$ is similarly higher than that for NGC 3115 (see Table 1). In the case of NGC 1600, however, its mass is higher than that of M87 by a factor of ~3, but its $\dot{M}_B$ is essentially the same as that of M87. Be as it may, the derived $\dot{M}_i$ in all 3 cases are either essentially identical or within a factor of < 2 of the reported $\dot{M}_B$. It is noteworthy that in comparison to similar-size BHs at z >5.7 (see Table1 in A23), the derived $\dot{M}$ of these 3 BHs are 2-3 orders of magnitude smaller and radiative efficiencies higher by factors of ~4-5.

Table 1: Predicted and known properties of nearby SMBHs

| Black Hole | z | Log $M_{BH}$ ($M_\odot$) | ε | $\dot{M}$ ($M_\odot$/yr) | $\dot{M}_i$ ($M_\odot$/yr) | $\dot{M}_B$ ($M_\odot$/yr) | $L_{bol}$ x$10^{45}$ ergs | Ref |
|---|---|---|---|---|---|---|---|---|
| Messier 87 | 0.00428 | 9.48-9.82 | 0.84 | 0.015-0.033 | 0.1-0.21 | 0.1-0.20 | 4.5-9.8 | 1,2 |
| NGC 3115 | 0.0022 | 9.0 -9.3 | 0.84 | 0.005-0.01 | 0.031-0.062 | 0.02-0.04 | 1.5-3 | 3,4 |
| NGC 1600 | 0.0156 | 10.23±0.04 | 0.68 | 0.088±0.007 | 0.275±0.025 | 0.15 ± 0.05 | 10.6 ± 0.9 | 5 |
| Sgr A* | 0.00002 | 6.618 | 0.90 | 2x$10^{-5}$ | 2.x$10^{-4}$ | > 5x$10^{-5}$<br>1.6-3.2x$10^{-4}$ | 0.0013 | 6<br>7 |
| Cygnus A | 0.056 | 9.4 ±0.12 | 0.87 | 0.015 ± 0.004 | 0.12 ± 0.03 | ≥ 0.15 | 5.7 ± 1.5 | 8 |

$M_{BH}$ = BH mass in solar masses ($M_\odot$)
$\dot{M}$ = Predicted mass accretion rate in solar masses $M_\odot$/year.
$\dot{M}_i$ = Predicted inflow rate $\dot{M}_i = \dot{M}/(1-\varepsilon)$, where ε is the inferred radiative efficiency.
$\dot{M}_B$ = Bondi accretion rate reported by works referred to in the last column.

$L_{bol} = (\dot{M}c^2 \varepsilon)/(1-\varepsilon)$, where c is velocity of light is the bolometric luminosity.

References: 1= Di Matteo et al. (2003); 2 = Russell et al. (2015); 3 &4 = Wong et al. (2011 & 2014); 5 = (Runge and Walker, 2021); 6 = Quataert, Narayan, & Reid (1999); 7= Cocker & Melia (1997); 8 = Lo et al. (2021)

For Sgr A*, QNR1999 obtained $\dot{M}_B$ ~3x$10^{-5}$ $M_\odot$/yr for $M_{BH}$ =2.6x$10^6$ $M_\odot$ using the observed distribution of nearby mass-losing stars and indicated that this estimate is a lower limit. Since the current best estimate of $M_{BH}$ for Sgr A* is higher by ~1.6, the value shown in Table 1 is scaled up by a factor of 1.6. The predicted $\dot{M}_i$ is within a factor of ~4 of the reported lower limit of $\dot{M}_B$. And CM1997 obtained a Bondi capture rate of 1-2x$10^{-4}$ $M_\odot$/yr using Bondi-Hoyle 3D hydro dynamical numerical simulations of accreting winds from nearby massive stars. Scaling their estimate up also by a factor of 1.6, one gets a Bondi capture rate of ~ 1.6-3.2x$10^{-4}$ $M_\odot$/yr shown in Table1 in remarkably good agreement with the predicted $\dot{M}_i$. Besides these two estimates based on Bondi-Hoyle accretion theory, QNR1999 obtained an upper limit of ~8x$10^{-5}$ $M_\odot$/yr (scaled-up to ~1.3x$10^{-4}$ $M_\odot$/yr) using X-ray data and infrared emissions from Sgr A*; and Cocker and Melia (2000) obtained a mass-inflow rate of ~$10^{-4}$ $M_\odot$/yr (scaled up to 1.6x$10^{-4}$ $M_\odot$/yr) using a spherical accretion model. Both estimates are in good agreement with the predicted $\dot{M}_i$. Taking, however, an average of these four estimates of $\dot{M}_i$ not based of any of the hot accretion models, we get $\dot{M}_i$ =1.45x$10^{-4}$ $M_\odot$/yr including the lower limit and $\dot{M}_B$ ~1.8x10$^{-4}$ M$_\odot$/yr excluding the lower limit. Both of these averages are in excellent agreement with the predicted $\dot{M}_i$ ~2x10$^{-4}$ M$_\odot$/yr. In contrast ADAF models give mass-inflow rates <10$^{-5}$ M$_\odot$/yr. (Narayan et al., 1998; Quataert and Narayan, 1999), and ADIOS models require accretion rates ~ 2 or more orders of magnitude smaller (Blandford and Begelman, 1999).

Lo et al. (2021) used polarimetric observations and assumptions concerning Faraday rotation measure to constrain the Bondi accretion rate of Cygnus A as a function of gas density. The lower limit of $\dot{M}_B$ =0.15 M$_\odot$/yr shown in Table 1 corresponds to a density profile with a power-law index (PLI) of 3/2 and the upper limit of 2.25 M$_\odot$/yr (not shown in Table 1) corresponds to a PLI of 1/2. The predicted $\dot{M}_i$ concurs with the lower limit of $\dot{M}_B$ taking into consideration the uncertainty in $\dot{M}_i$ resulting from the uncertainty in $M_{BH}$. The upper limit of $\dot{M}_B$ is, however, more than an order of magnitude greater than $\dot{M}_i$. We can discriminate between the two extremes using the "observed" and predicted bolometric luminosity $L_{bol}$. Vasudevan et al. (2010) determined $L_{bol}$ =5x10$^{45}$ ergs/s for Cygnus A based on X-ray data and infrared emissions. We get $L_{bol}$ = (5.7$\pm$1.5) x10$^{45}$ ergs/s (see Table 1). The predicted and observed $L_{bol}$ agree remarkably well. This agreement between the predicted and observed $L_{bol}$ combined with the agreement between $\dot{M}_i$ and the lower limit of $\dot{M}_B$, suggests that $\dot{M}_B$ of Cygnus A is at or close to the lower limit shown in Table 1 corresponding to a density PLI of ~3/2.

The foregoing comparisons and analyses indicate that there is no significant difference between the reported Bondi accretion rate and the predicted mass inflow rate onto the BH, not in one or two but in all five cases. It is highly unlikely that this finding is simply fortuitous. Hence, we can safely conclude that the mass inflow rate from the Bondi radius to the BH remains essentially constant as envisioned in the ADAF model (e.g. Narayan and Yi, 1994), but exclude the ADIOS and CDAF models that envision a decrease in $\dot{M}_B$ (Blandford & Begelman, 1999, 2004; Narayan et al. 2000). If hypothetically, however, the accretion rate were to decline proportional to say the square-root of the distance to the BH, then as per the calculations of NY14 as little as 0.1-.03% of $\dot{M}_B$ would make it to the BH. The results not only clearly exclude such a possibility, but show that the 5 BHs accrete at a rate (~10-32% of $\dot{M}_B$) that is an order of magnitude higher irrespective of how much of $\dot{M}_B$ reaches the BH. The classic Bondi/ADAF model of hot accretion flow, however, predicts a density profile at the Bondi radius with a power-law index of 3/2, but the available density profiles analyzed later show mixed results and seem to support a somewhat flatter power-law index close to 1.

The inferred radiative efficiencies $\varepsilon$ of the 5 BHs provide additional insights into the accretion process and constraints for accretion models. Sgr A* is the smallest of the 5 BHs, and the results show that its accretion rate $\dot{M}$ is the lowest, and its radiative efficiency $\varepsilon$ is the highest. At the other end of the spectrum, NGC 1600 is the largest, its $\dot{M}$ is the highest, and its $\varepsilon$ is the lowest. And M87, NGC 3115, and Cygnus A have similar masses, their $\dot{M}$ are similar, and their $\varepsilon$ are also similar. The inferred $\varepsilon$ are consistent with the finding by A23 that $\varepsilon$ is an inverse function of BH mass $M_{BH}$ for BHs at similar redshifts. The inferred $\varepsilon$ for the 5 BHs are ~7-9 times higher than the commonly used value of 0.1, and orders of magnitude higher than that expected from hot accretion flow models of LLAGNs such as Sagittarius A* (see review by YN14). In fact, these 5 SMBHs have much higher $\varepsilon$ than similar-size highly luminous BHs at z >5.7 (see Table 1 in A23), but for some reason the dissipated energy is not reflected in their observed luminosities except for Cygnus A. The "observed" and predicted $L_{bol}$ for Cygnus A match surprisingly well as indicated above. For the other four BHs, the predicted $L_{bol}$ are orders of magnitude higher than the observed X-ray and or nuclear luminosities (see R15; W14, RW21; QNR 1999).

In the ADAF model of hot accretion flow most of the dissipated energy is advected; which might explain why the observed luminosities are so low; but it is enigmatic why the observed luminosity of Cygnus A is comparable to its predicted bolometric luminosity $L_{bol}$ while the observed luminosities of the other 4 BHs are lower than their predicted $L_{bol}$ by orders of magnitude. Moreover, in hot accretion flows, the radiative efficiency apparently decreases with decreasing mass accretion rate (see YN14), contrary to the results of this study and findings in A23. As pointed out earlier, Sgr A* has the lowest $\dot{M}$ but its ε is the highest of the 5 BHs. And NGC 1600 has the highest $\dot{M}$, and its ε is the lowest. Also, A23 showed that at lower redshifts as a BH ages, its accretion rate decreases but its radiative efficiency increases. It is beyond the scope of this paper to even attempt to reconcile these apparent contradictions between the theory behind ADAF model of hot accretion flow and the findings of this letter based on empirical relations.

## 4. PHYSICAL BASIS

The foregoing agreement between $\dot{M}_i$ and $\dot{M}_B$ is, a priori, surprising for 2 reasons. First, the empirical relations 1 and 2 are based on the data of the youngest highly luminous quasars, whereas the 5 BHs are separated in time by ~13 billion years from the high-z quasars and apparently belong to the class of LLAGNs. Second, the empirically derived $\dot{M}_i$ is based on the Salpeter relation (Salpeter, 1964) and depends only on a BH's $M_{BH}$ and z (see Eq.1), whereas $\dot{M}_B$ based on Bondi accretion theory is a function of temperature and density in addition to $M_{BH}$. It is instructive to deconstruct and compare the forms of $\dot{M}_i$ and $\dot{M}_B$. Since the ambient matter/gas density ρ in the interstellar medium at any z to that at z=0 scales as $(1 + z)^3$, the empirically derived $\dot{M}_i \propto \rho M_{BH}$ for a given radiative efficiency. On the other hand, $\dot{M}_B \propto c_s^{-1} \rho_B R_B M_{BH}$, where $\rho_B$ and $c_s$ are gas density and sound speed at the Bondi radius $R_B$. The sound speed $c_s \propto T_B^{1/2}$, where $T_B$ is the temperature at the Bondi radius. Theoretically, T decreases away from the BH following a power-law index. Let us assign a power-law index p such that $T \propto R^{-p}$, where p has an unknown value between 0 and 1. Equating $\dot{M}_i$ and $\dot{M}_B$ for a given radiative efficiency, we get:

$$\rho M_{BH} \propto \rho_B R_B R_B^{0.5p} M_{BH} = \rho_B R_B^{(1+0.5p)} M_{BH} \qquad (3)$$

From which, we get:

$$\rho_B \propto (\rho R_B^{-(1+0.5p)}) \qquad (4)$$

Equation 4 implies that density ρ(R) is a function of temperature and that ρ(R) at the Bondi radius is $\propto R_B^{-(1+0.5p)}$, where p = 0-1 is the power law index for temperature at the Bondi radius. Hence, equating the empirically derived $\dot{M}_i$ to $\dot{M}_B$ indicates that the density power-law index (PLI) should be $\geq 1$ but $\leq 1.5$ with the actual value dependent upon the temperature PLI. Let us examine the available data for density profiles.
.
For M87, R15 did not find any significant increase in temperature from the Bondi radius towards the BH, and determined a PLI = 1-1.2 ± 0.2 for density that is in accord with a PLI =1 predicted by Eq.4. More recently, R18 found variations in density gradient along different directions perpendicular to the jet axis of M87; the mean value of the power-law indices is, however, consistent with a PLI ~1. For NGC 3115, W11 determined a PLI =1.03 $\pm$ 0.2 for density, and more recently W14 obtained a temperature PLI of p ~0.445 and several density profiles depending upon the temperature model used. The PLI indices of these profiles near the Bondi radius are in agreement with the earlier determination, but those closer to the BH seem to yield flatter PLI. Using the p value or the PLI index for temperature obtained by W14, we get from Eq.4 a predicted density PLI ~1.22 for NGC 3115 that agrees with its reported PLI =1.03 $\pm$ 0.2 near the Bondi radius. In contrast, RW21 obtained essentially a flat temperature PLI p=0.073 for NGC 1600 and a density PLI =0.61+0.13 that is ~ 25-40% lower than the PLI of ~1 predicted by Eq.4. We note, however, that there are only 2 data points that define the density PLI of NGC 1600 inside its Bondi radius; and given the uncertainties in them a density PLI of ~1 cannot be ruled out (see Fig.3 in RW21). For Cygnus A we concluded above that its $\dot{M}_B$ based on a density PLI of 1.5 is the most likely. The density profile of Sgr A* is apparently not known; although citing Wang et al. (2013), YN14 observed that Chandra observations of Sgr A* indicate a flat radial density profile with a PLI index of 1 near the Bondi radius. Thus, although the available density profiles show mixed results, the bulk of available data (possibly 4 out of 5 cases) are in accord with the predictions of Eq.4. More observational data are needed to ascertain whether there is a power-law index for density at the Bondi radius common to most if not all LLAGNs. Equation 4, however, indicates that there need not be a density PLI common to all BHs but may vary within the limits of 1-1.5 depending upon the temperature profile.

## 5. CONCLUSIONS

We determined the accretion rate $\dot{M}$ and radiative efficiency $\varepsilon$ for the SMBHs in our and nearby galaxies M87, NGC 3115, NGC 1600, and Cygnus A using simple well-tested empirical relations defined by Eq.1 and 2. The results are shown in Table 1. The principal findings and conclusions are as follows.

1. In each case, the predicted mass-inflow rate $\dot{M}_i = \dot{M}/(1-\varepsilon)$ onto the BH is remarkably similar to the reported Bondi accretion rate $\dot{M}_B$, indicating that there is no significant discernable change in $\dot{M}_B$ from the Bondi radius to the SMBH (see Table 1).
2. The results do not support the ADIOS and CDAF variants of hot accretion flow in which the mass-inflow rate decreases from the Bondi radius to the BH.
3. The predicted accretion rates $\dot{M}$ of the 5 BHs are lower by 2-3 orders of magnitude and their $\varepsilon$ higher by factors of 4-5 than for their counterparts or similar-size BHs at higher redshifts at z>5.7 (for comparison see Table 1 in A23).
4. Sgr A* has the lowest accretion rate but its $\varepsilon$ is the highest of the 5 BHs, and NGC 1600 has the highest accretion rate and its $\varepsilon$ is the lowest – a finding contrary to the expectation in hot accretion flows that $\varepsilon$ decreases with decreasing accretion rate.
5. Approximately 10-32% of the mass available at the Bondi radius is accreted by the BHs, or ~68 to 90% is dissipated or radiated away.
6. For Cygnus A, the dissipated energy is reflected in its observed luminosity $L_{bol}$. In all other cases, the observed luminosity is orders of magnitude lower than $L_{bol}$ as observed by previous workers. This difference in the observed luminosities remains enigmatic.
7. In the ADAF model most of the dissipated energy is advected; which might explain why the observed luminosities of 4 BHs are so low, but available density profiles seem to support a power-law index close to 1 rather than 1.5 in the classic Bondi/ADAF model. Moreover, the model needs to reconcile the finding enumerated as #4 with the theory behind hot accretion flows.
8. The equivalence of the empirically derived Eq.1 with Bondi accretion predicts that the density power-law index at the Bondi radius should be $\geq 1$ and $\leq 1.5$. The available density profiles show mixed results, but the bulk of the available data support such a prediction.

**DATA AVAILABILITY**: *No new data were generated or analyzed in support of this research.*